\documentclass[noacm]{acmart}
\renewcommand\footnotetextcopyrightpermission[1]{}

\AtBeginDocument{%
  \providecommand\BibTeX{{%
    \normalfont B\kern-0.5em{\scshape i\kern-0.25em b}\kern-0.8em\TeX}}}

\setcopyright{none}

\usepackage{graphicx}

\usepackage{multirow}
\usepackage{soul}
\usepackage{dirtytalk}

\usepackage{subcaption}
\usepackage{enumitem}
\usepackage{multirow}
\usepackage{color}
\usepackage{cleveref}
\usepackage{booktabs}

\usepackage{graphicx}
\usepackage{subcaption}

\usepackage{makecell}

\crefname{equation}{Eq.}{Eq.}
\crefname{section}{Section}{Sections}
\crefname{subsection}{Section}{Sections}
\crefname{subsubsection}{Section}{Sections}
\crefname{figure}{Figure}{Figures}
\crefname{table}{Table}{Tables}
\crefname{subfigure}{Figure}{Figures}
\crefname{algocf}{Algorithm}{Algorithms}

\newcommand{\xhdr}[1]{\vspace{0.05in}\noindent{{\bf #1}}}

\setcopyright{none}
\begin{document}

\title{Corporate Communication Companion (CCC): An LLM-empowered Writing Assistant for Workplace Social Media}

\author{Zhuoran Lu}
\authornote{Purdue University, West Lafayette, Indiana, United States. Email: lu800@purdue.edu. Work done while working at Microsoft.}

\author{Sheshera Mysore}
\authornote{University of Massachusetts Amherst, Amherst, Massachusetts, United States. Work done while working at Microsoft.}

\author{Tara Safavi, Jennifer Neville, Longqi Yang, Mengting Wan}
\authornote{Microsoft, Redmond, Washington, United States. Email: mengting.wan@microsoft.com. \\Some of the information in this document relates to pre-released content which may be subsequently modified. Microsoft makes no warranties, express or implied, with respect to the information provided here. This document is provided “as-is”. Information and views expressed in this document, including URL and other Internet Web site references, may change without notice. Some examples depicted herein are provided for illustration only and are fictitious. No real association or connection is intended or should be inferred. This document does not provide you with any legal rights to any intellectual property in any Microsoft product.© 2024 Microsoft. All rights reserved.}

\renewcommand{\shortauthors}{}
\renewcommand{\shorttitle}{Corporate Communication Companion (CCC)}

\begin{abstract}
Workplace social media platforms enable employees to cultivate their professional image and connect with colleagues in a semi-formal environment. While semi-formal corporate communication poses a unique set of challenges, large language models (LLMs) have shown great promise in helping users draft and edit their social media posts. However, LLMs may fail to capture individualized tones and voices in such workplace use cases, as they often generate text using a ``one-size-fits-all'' approach that can be perceived as generic and bland. In this paper, we present Corporate Communication Companion (CCC), an LLM-empowered interactive system that helps people compose customized and individualized workplace social media posts. Using need-finding interviews to motivate our system design, CCC decomposes the writing process into two core functions, outline and edit: First, it suggests post outlines based on users’ job status and previous posts, and next provides edits with attributions that users can contextually customize. We conducted a within-subjects user study asking participants both to write posts and evaluate posts written by others. The results show that CCC enhances users’ writing experience, and audience members rate CCC-enhanced posts as higher quality than posts written using a non-customized writing assistant.  We conclude by discussing the implications of LLM-empowered corporate communication.
\end{abstract}

\received{20 February 2007}
\received[revised]{12 March 2009}
\received[accepted]{5 June 2009}

\maketitle

\section{Introduction}
Workplace social media platforms enable people to continuously cultivate their professional images by both showcasing their personal influence and forging connections within the organization ~\cite{damnianovic2012role,sun2021impression}. 
In this way, such platforms have a unique impact on corporate communication that goes beyond conventional channels like emails, reports, etc \cite{bertin2020collaboration,kane2015enterprise,brzozowski2009watercooler}. However, crafting a consistent and clear professional image is a challenging and long-term endeavor \cite{sun2021dark,huang2015structural}. 
It requires people to align their behavior with specific strategies and create content they believe will be advantageous to their social media image \cite{yang2022social,sun2021impression}. 

Given that different people employ different strategies to manage their profiles~\cite{johnson2017importance,pan2017understanding,saha2019libra}, it becomes even more challenging when one's desired workplace image differs from the communication style to which they are accustomed~\cite{van2013you,bremner2015learning}. Consider Alice, a senior engineer who is used to writing technical documents but now finds herself in an unfamiliar scenario of crafting a post to invite interns to a casual reception. Users like Alice must consider multiple, potentially conflicting goals in post-writing to best serve her professional image: How can she make the post personal enough to avoid the perception of being templated, but also appropriate to be published in the workplace? How can she write a post that is both concise but informative? Such a predicament highlights the challenges faced by users like Alice when they try to tailor their behavior on workplace social media to enhance their professional image. This process can be time-consuming and may discourage people from actively participating on workplace social media, which could ultimately undermine knowledge sharing and community dynamics within the corporate environment \cite{chen2017innovation,sharma2022making}.

The advancement of artificial intelligence (AI) technologies, particularly large language models (LLMs), have shown promise in addressing this issue by facilitating collaborative writing in various communication contexts \cite{chen2019gmail,kim2023towards,yuan2022wordcraft}. For example, prior research has demonstrated the potential for human-AI co-creation of presentation slides from coding notebooks \cite{zheng2022telling}, highlighting AI's ability to effectively mediate well-defined, formal corporate communication. However, workplace social media poses its unique challenges. Alice's example is just one of many scenarios that workplace social media users encounter. Content generation on workplace social media involves a wide range of topics, each requiring different levels of tailoring under different personal preferences. Even with the help of LLMs, users still need a clear mental model to guide how they should interact with LLMs through prompts and eventually produce satisfactory content. Unfortunately, it is often not the case, as end-users tend to prompt LLMs opportunistically, without providing sufficient contextual information and systematical tailoring details \cite{zamfirescu2023johnny}. Specifically in workplace social media writing, failure to do so could not only harm the content quality but also damage their workplace image. 

Given these challenges, in this study, we set out to design a novel interactive tool to support collaborative writing with efficient, controllable, and systematic tailoring of posts on workplace social media. Recognizing the unique affordances of workplace social media as revealed by previous literature, we sought to understand users' specific needs for writing assistance in this context, in order to design a tool that could enhance both their writing experience and post quality. To answer this question, we conducted a two-phase study on a workplace social media platform within a major organization.

To gain an initial understanding of how people write workplace social media content, we performed an offline data analysis before conducting our formal two-phase study. We applied text clustering methods to a real-world workplace social media dataset and examined the relationships between the topics and the writing styles of these social media posts. The results confirmed our hypothesis that people adopted vastly different writing strategies to suit their personal preferences and situational contexts when communicating about even the same topics.

Building on the insights from the offline data analysis, in Phase 1, we conducted a series of theory-informed, semi-structured interviews with 7 active workplace social media users. These interviews helped us to define the design goals of our Corporate Communication Companion (CCC) system and allowed us to iteratively implement it. 
Specifically based on our findings from the interviews, we identified different user scenarios and classified them into two categories: (1) Work-related posts, which requires a \textbf{professional} style; (2) Leisure-related, which requires a \textbf{casual} tone. 
These categories helped us create two types of user profiles for each user based on their historical posts and job status, thus enabling us to provide personalized writing assistance by recognizing user intentions and their style preference.
We further decomposed the writing process into two stages, Co-Outline and Co-Edit, and provide assistance through CCC for each stage. In the \textbf{Co-Outline} stage, we leveraged the user's job status and prior outlines to generate backbone ideas that can serve as the scaffolding for the current post development. In the \textbf{Co-Edit} stage, users can customize various attributes related to the tone and style of their posts, allowing the CCC to generate desirable transformations from drafts, keywords, or instructions from users.

In Phase 2, to evaluate the usability and effectiveness of CCC, we conducted a within-subjects user study with 10 participants from a large corporation. We created a comparison
baseline by disabling the customization and writing process decomposition functions of CCC. This baseline reflects a standard LLM writing assistant (e.g., GPT-like product). The user study consisted of two parts conducted on two separate days. On Day 1, participants wrote posts using both systems, and on Day 2, they evaluated posts composed by other participants. The results show that CCC benefits workplace social media communication from the perspective of both \textbf{writers} and \textbf{audience members}. For writers, CCC enhances the writing experience by fostering a sense of collaboration and engagement in the writing process. Furthermore, posts generated with CCC were perceived as more complete and unique, leading to higher levels of trust and a sense of writing success. Importantly, these improvements in user experience do not require an increased cognitive load for users. From the audience's standpoint, participants consistently favored posts authored with CCC assistance, rating them as more informative, engaging, and appropriate.

Our findings make unique contributions to the human-computer interaction field, especially human-AI interaction and computer-supported cooperative work, in several ways. First, we proposed novel LLM-empowered writing assistant designs to help users write workplace social media posts and establish their professional images. Our user study revealed both the benefits and the potential limitations of LLM-empowered writing assistants. Second, we addressed the challenges of using LLMs to assist semi-formal corporate communication. 
We enhanced people's writing initiation by providing contextual information and decomposing the writing process into multiple steps. Furthermore, we enabled people to customize the assistant to fit their personal writing styles and to tailor the content generation more efficiently and transparently. Together, these features contributed to a mixed-initiative interaction between people and the assistant, which improved both the quality and the perception of the posts. Finally, our results sparked design implications for LLM-empowered writing assistants for broader communication scenarios in the future.   


\section{Related Work}

\subsection{Social Media in Workplace}
\label{workplace_social_media}
Workplace social media, also known as enterprise social media, has rapidly emerged as a necessity in modern corporations \cite{leonardi2013enterprise,kane2015enterprise,van2015enterprise}. Platforms such as LinkedIn, Yammer, and Slack have been proven beneficial for professional connections \cite{dimicco2008motivations}, knowledge sharing \cite{ellison2015use,sun2019impact}, improvement of productivity  \cite{aguenza2012social}, and establishment of users' personal and corporate brands  \cite{whitmer2019you,sun2021impression}. In addition to such professionally developed workplace social media, some specific communities on conventional social media now also serve as actual enterprise social media. For instance, one previous study~\cite{collins2016scientists} revealed scientific communication on social media and its potential benefits. Academic usage of Twitter is another case of mixed usage between personal and workplace social media \cite{mohammadi2018academic}. 

Unlike personal interactions on conventional social media or professional communication through reports and emails, workplace social media is built for a blend of different types of communication, brand promotion, and organizational culture cultivation within the corporation \cite{liu2019enterprise,yee2021and}. As a result, people are motivated by different incentives, leading to a wide range of behavioral patterns on their workplace or work-related social media. To begin with, there exist typical types of dual behaviors, including productive-behavior and unproductive-behaviors ~\cite{carlson2016social}. For instance, users create posts about personal experiences for social integrative and entertainment gratification on Yammer~\cite{zhong2017employees}. Meanwhile, senior management values more about the information sharing and broadcasting function of Yammer, prioritizing its role as productive communication~\cite{hall2015analysing}. Such variety extends in different levels of  communication~\cite{riemer2012oh}. At the linguistic level, people adopt different styles. Sjölund ~\cite{sjolund2016discursive} identified five key discourses among Yammer data, revealing the presence of distinct tones and language styles. For instance, Conversation enablers express themselves in a personal and expressive manner, whereas the Communal entity employs expressive language with an emphasis on togetherness.
Beyond the linguistics, users' communication structure depends on the audience that they are targeting as well: a case study on Yammer revealed that net-wide communication visible to all the company and local community engagement are different~\cite{riemer2013role}. However, there remains a gap in our understanding, as there has yet to be a comprehensive and detailed exploration of the landscape of user-generated content on workplace social media across various use cases and scenarios.

\subsection{Human-AI Co-creation}
In recent years, the advancement of AI technologies has significantly enhanced the collaboration between humans and AI~\cite{amershi2019guidelines, glikson2020human}, particularly in the domain of content creation and writing~\cite{dang2022beyond,soni2018systematic}. Historically, intelligent writing assistants were mainly geared towards grammar checks to ensure clarity in writing \cite{soni2018systematic}. Tools such as Grammarly have become widely used in English writing practices \cite{ghufron2018role}. However, with the development of deep learning technologies, the role of AI in content creation has evolved from simple editing and double-checking \cite{summerville2018procedural}. Modern generative machine learning models have endowed AI writing assistants with a higher level of competence, fostering a synergistic relationship with human writers, where humans often find themselves learning from their AI collaborators \cite{pavlik2023collaborating}. The range of applications spans from general creative aims like script writing \cite{biermann2022tool,clark2018creative} to more structured forms of writing, including essays \cite{fitria2023artificial} and scientific articles \cite{salvagno2023can}. Moreover, the human-AI co-creation paradigm has been generalized to more than writing. More modalities are now incorporated in this co-creation synergy. For instance, \cite{louie2020novice} designed an AI-based tool for music creation; \cite{lawton2023tool} evaluated people's perceptions in the co-drawing process with AI.

Further empirical studies probed into evaluating both performance and people's perceptions or experience in collaborative writing with AI \cite{shen2023parachute,lee2022coauthor}, in order to better collaborative composition systems. \cite{jakesch2023co,mirowski2023co} found that opinionate language models can lead to writers' opinion shifts and students may use it to cheat \cite{fyfe2022cheat}, which indicate the potential risks of collaborative writing. At the same time, previous research was also approaching to personal preferences in such collaborative writing. As revealed in \cite{biermann2022tool}, writers want AI to respect their values and strategies.

Despite all these progressions in human-AI co-creation, a set of key components that has been a longstanding focus in conventional intelligent systems, namely customization and personalization, remains absent in the current landscape. Technologies ranging from collaborative filtering~\cite{konstan1997grouplens,su2009survey,he2017neural} to graph neural networks ~\cite{gao2023survey, gasteiger2018predict} actively strive to enable AI-based systems to cater to people's specific preferences, recognizing that individuals present distinct needs and requirements when seeking assistance from intelligence systems. Creation, given its inherently subjective nature involving human intelligence~\cite{scott1999agency,emig1977writing}, is particularly ripe for the integration of personalization. Pioneering efforts are emerging to bridge the gap between LLMs and personalization. For instance, LaMP~\cite{salemi2023lamp} proposed a benchmark designed to train and assess LLMs in generating personalized outputs. While such work has made strides in this direction, it has generally revolved around the concept of tailoring outputs at a relatively coarse level based on user preferences. However, there has been a notable absence of research dedicated to exploring human-LLM co-creation, leveraging users' information to enhance the co-creation process.

\subsection{LLM-Assisted Content Creation}
In the expanding field of human-AI co-creation, LLMs, such as GPT-3 \cite{brown2005language}, LaMDA \cite{thoppilan2022lamda}, ChatGPT \cite{openai2023chatgpt},  have recently started their crucial and evolving roles in aiding content generation. Their integration into a wide range of applications, including but not limited to story writing \cite{yuan2022wordcraft}, lyric composition \cite{crothers2023bloom}, and coding \cite{liu2023your,cai2023low}, has become increasingly prevalent. For instance, \cite{chew2023llm} found that the GPT 3.5 could generate deductive coding results at levels of agreement comparable to human coders, indicating LLM's potential in open-end content generation for academic purpose. In addition, as a complementary to human creators, LLMs could provide help to improve human-generated content as well. As a prevalent example, LLM-empowered content moderation \cite{kumar2023watch} shows the potential to remind human writers of the language use and lead to more harmonious content generation. As discussed in \cite{gmeiner2023dimensions}, LLMs could be capable of serving as \textit{design materials} in supporting a wide range of language-based content-generation to enhance both productivity and creativity. 

Despite the advancing capabilities of LLMs, LLM-assisted content creation still face significant challenges. To begin with, LLMs themselves particularly lack in capturing full contextual information and the subtleties lying in human language \cite{binz2023using,santhanam2019emotional}. Issues such as misinterpretation of nuances, failure to capture deeper meanings or sentiments \cite{wang2023emotional}, and difficulties in maintaining context across larger text windows \cite{xu2023retrieval,xiong2023effective} have been commonly detected by recent studies. Moreover, a more severe problem here presented lies in the human-LLM collaboration process. Recent research has explored LLM's role serving as initial draft provider, content editor, etc \cite{lingard2023writing,macdonald2023can, kim2023using}. However, these studies often consider these roles in isolation. There is a lack of comprehensive discussion about the specific roles LLMs should play and the unique interaction mechanisms they should adopt in collaboration with human creators \cite{subramonyam2023bridging}. While LLMs are capable of generating human-like text, the potential of their assistance extends beyond that of traditional human assistants. In previous studies in designing human-LLM collaboration, \cite{moore2023empowering, li2023can} presented novel interfaces to enhance human-LLM collaboration via more advanced LLM-assistance. However, the efficient utilization of such assistance from a human-centric perspective remains scarce. Key questions include: How can we conceptualize, decompose, and design the human-LLM collaboration process for content generation? How can we adapt a general LLM to specific subdomains of collaboration by leveraging user information? Such questions are crucial while under-explored in LLM-assisted content creation.




\section{Study Overview}

In this study, we aim to present a comprehensive design procedure for an intelligent writing assistant to support users like Alice on workplace social media. To achieve this goal, we started by the offline data analysis then conducted a two-phase study targeting different research questions in each phase.

We start by anchoring query of the entire study: \textbf{how do people write posts on workplace social media?} To answer this question, we applied clustering analysis on a set of posts collected from a workplace social media platform used in a large corporation. The results confirmed that beyond the diverse use cases and writing styles on workplace social media, people tend to have different strategies for cultivating their digital identity, thus leading to a blending of topics and writing styles. 

Proceeding from the point, in the first phase, we want to understand \textbf{how LLM-empowered writing assistant could help users write workplace social media posts.} Therefore, we began with a series of semi-structured needs-finding interviews. From the interviews, we identified two typical use cases of workplace social media: work-related and leisure-related posts. Furthermore, we leveraged two sets of theory-informed questions to probe the potential utilization of LLM for users' writing. The first set focused on how users craft their workplace social media posts on their own in the three stages of writing \cite{flower1984images,flower1981cognitive}. The second set of questions presented eight features that corresponded to different affordances of workplace social media \cite{sun2019impact,sun2021impression} (e.g., help users maintain a consistent language style for the persistence of their posts), and tested how users responded to these proposed features. At the same time, we developed an initial prototype system Corporate Communication Companion (CCC); we iteratively tested and improved its usability with participants, which enabled us to finalize its design. 

Finally, in the second phase, we examined  \textbf{how the proposed assistant helped users write on workplace social media.} To assess the effectiveness of the proposed system, we conducted a within-subject user study with 10 participants. In the study, participants wrote with CCC on both work-related and leisure-related tasks, and then evaluated the posts written by others, thus enabling us to evaluate CCC from both the writers' and audience members' perspective. We found that CCC enhanced users' writing experience by providing customization and efficient interaction with LLMs, and ultimately improved the writing quality perceived by the audience. We also identified the pitfalls of the current CCC system and discussed the implications of our findings.

\section{Offline Data Analysis}
\label{offline_data_analysis}
As a starting point of our study, we would like to embark on this question: what do people write about, and how do they write on workplace social media? Answering this question helps us to position our study in the specific setting of workplace social media that differs from a writing scenario in other online communities.

Proceeding from this standpoint, we first collected a representative offline dataset from a real workplace social media platform used in a large corporation. Then, we performed a clustering analysis to investigate the taxonomy of users' typical use cases. In particular, we are interested in the following two dimensions of user posts: \textbf{topic} and \textbf{style}. 


\begin{figure*}[t]
\centering
\includegraphics[width=1.0\textwidth]{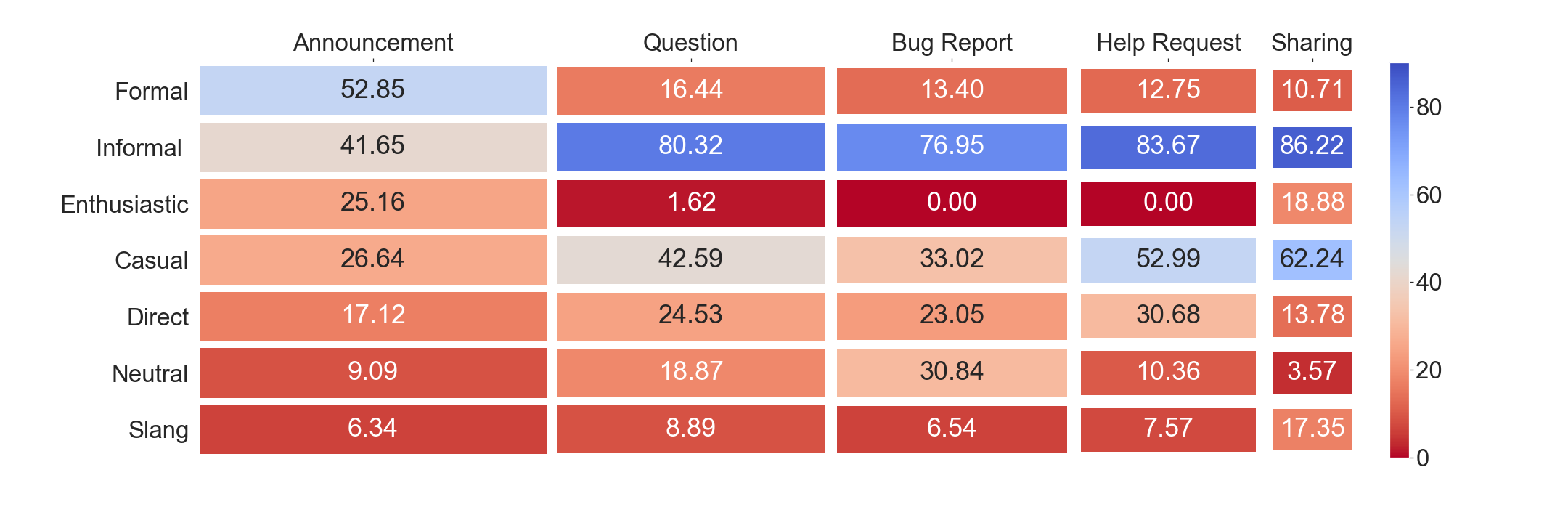}
\vspace{-10pt}
\caption{A visualization of users' post writing varying along \textbf{Use Cases} and \textbf{Writing Style}. Each column represents posts related to a specific topic, and the width of each column corresponds to the proportion of that topic among all posts. For example, the first column represents posts about announcements, which is the most frequently written topic among all posts. The numbers within each cell indicate the percentage of a specific writing style within a particular topic. For instance, the cell in the upper left corner shows that $52.85\%$ of the announcement posts have a formal tone. Note that the clustered styles are not mutually exclusive; for example, an announcement post can be both enthusiastic and use slang in writing. Hence, the sum of percentages within each column does not necessarily add up to $100\%$.}
~\label{fig:distribution}
\vspace{-10pt}
\end{figure*}

Our dataset consists of posts authored on the workplace social media platform within a two-week time window. After filtering out inattentive posts with less than three sentences, we randomly sampled 2,000 posts for our clustering analysis. We leveraged the capabilities of Large Language Models (LLMs) to conduct a step-wise clustering. The analysis procedure is described below:
\begin{itemize}
    \item We first prompted GPT-4 \cite{peng2023instruction} to perform two tasks: (a) Summarize each post within three sentences, and (b) Describe the writing style of each post within three sentences.
    \item To generate the post type cluster labels, we randomly sampled 200 post summaries and prompted GPT-4 to generate the 10 most frequent post types observed in the data (e.g., ``announcement'', ``bug report''). 
    \item We then prompted GPT-4 to assign each post a label from the 10 categories derived from the previous step. 
    \item We manually developed a set of keyword-based labels from the writing style summaries (e.g., ``formal'', ``informal'', ``slang''), and adopted a similar process to assign these labels to each post.
\end{itemize}

\cref{fig:distribution} shows the results of the clustering. In this figure, each column represents posts related to a specific topic, and the width of each column corresponds to the proportion of that topic among all posts. For example, the first column represents posts about announcements, which is the most frequently written topic among all posts. The numbers within each cell indicate the percentage of a specific writing style within a particular topic. For instance, the cell in the upper left corner shows that 52.85\% of the announcement posts have a formal tone. \footnote{Note that the clustered styles are not mutually exclusive; for example, an announcement post can be both enthusiastic and use slang in writing. Hence, the sum of percentages within each column does not necessarily add up to 100\%.}

By comparing the width of columns in \cref{fig:distribution}, we can see that topics covered on the workplace social media platform are mixed. Though the platform is populated with work-related, professional topics, it also incorporates a considerable amount of leisure-related, casual topics; thus confirming the previous findings that use cases on workplace social media are usually blended \cite{van2016classifying, el2015employees}. By taking a deeper look into the writing styles developed for each topic, we observe a crossover between formal and informal topics versus professional and casual writing styles. Taking announcements as an example, posts written in formal and informal styles are almost evenly distributed. It is intuitive to write a post in a professional tone, as for some users their writing styles align perfectly with the topics of their posts. However, interestingly, a casual writing style (26.64\%) and the use of slang (6.34\%) also make up a significant proportion of announcements. Vice versa, the right upper cells (e.g., 10.71\% of Sharing) indicate that some users believe that even when sharing their experiences, they should use a formal tone. 

The results highlight the variety of topics and styles covered on the platform, especially a blending across informality and formality. An intuitive gauge will stop here, considering users are just different. However, we would like to ask the following question here, are users' actual personas fully consistent with the image that they would like to cultivate on workplace social media? In other words, are people adopting the various writing styles across topics totally based on their own preference in daily writing? The answer is probably not. For instance, \cite{fieseler2015professional} found the tendencies of users to create their unique organizational identifications by combining both their private information and professional profiles. 

Thus, this mixed range may not merely be about personal leanings, but also a weight among various social factors. Furthermore, the combination employed by an individual user indicates this user's strategic way of shaping the image within the company. For instance, the users who use an informal and engaging tone in formal posts indicate that they still consider engagement to be an essential factor when making announcements on workplace social media, even above formality. However, the employment of formal tones in casual writing may be attributed to some users' perception that workplace social media still falls within the realm of corporate communication. Therefore, it would be better to always keep serious so that their colleagues would also perceive their account with only neat and professional content involved. 

Naturally, from a broader perspective, we would like to ask: What benefits to their personal image motivate their writing, and how do they decide the writing style they want at that moment?  These considerations are vital for designing the tool to support workplace social media writing. However, from the standpoint of the offline data analysis, it may be challenging for us to obtain definitive answers to these questions. Therefore, we are here initiating a two-phase study to address this gap, aiming to present a comprehensive design procedure from a user-centric view for our writing assistant, Corporate Communication Companion (CCC), specifically tailored for workplace social media.





\section{Phase 1: What Do Users Need from a Communication Companion?}


In Phase 1, we conducted a series of semi-structured interviews to understand the user needs for writing assistance on workplace social media. We then developed the design goals of our system from the interview findings. Finally, we present the system we designed and implemented and demonstrate how it reflects the design goals derived from our interviews.

\subsection{Interview Design}


Our data analysis demonstrates that people have various objectives and exhibit diverse writing styles when using workplace social media, even within the same use case. Moreover, the existing literature (\cref{workplace_social_media}) corroborates that people's usage of workplace social media is complex and multifaceted. As a result, this poses challenges for designing an LLM-empowered tool to facilitate users' writing on workplace social media platforms. Nevertheless, one thing is clear: such a system should go beyond a one-size-fits-all approach and ackowledge the preferences and requirements of each user. Therefore, we aim to develop a tool that can adapt to users' needs and allow them to tailor and customize their workplace social media posts.

To accomplish this goal, we conducted our interviews with two main objectives in mind. The first is to investigate features that could enable more effective online activities on workspace social media platforms, under the assumption that users have diverse needs and preferences for workplace social media content creation. The second objective is to identify and address common usability issues that such a system may encounter.




In pursuit of our first objective, we structured the first part of the interview based on two sets of theory-informed questions. 
Among various previous research~\cite{rohman1965pre,fitzgerald2012towards,greer2016introduction}, the widely acceptable Flower and Hayes Model conceptualized writing as a three-step process~\cite{flower1984images,flower1981cognitive}: planning, translation, and reviewing. Consequently, we inquired about (a) how people typically generate ideas and outlines for their posts, (b) how they craft language around the ideas and outlines, and (c) how they review and revise these posts for any possible language refinement. These questions revealed participants' independent writing practices, especially in the context of workplace social media. These findings enabled us to break down the entire collaborative writing process into steps and align it with independent human writing, allowing us to provide more specific assistance at each step of the writing process.

We also leveraged the previous organizational behavior literature~\cite{drahovsova2016benefits,holtzblatt2013evaluating,lehner2013organize} to inform our second objective: how to design customized LLM-empowered writing tools that could help users achieve their goals more efficiently through workplace social media. We adopted a previous framework that highlighted four affordances of workspace social media~\cite{sun2019impact,sun2021impression}: \textit{visibility, persistence, editability, and association}. This formed the foundation of our design proposal and helped us align the proposed features with user needs. We therefore compiled a list of attributes that our system could embody, such as supporting users' unique language styles, post structure, broadcasting, and more. Through a series of iterations, we carefully mapped each attribute to its potential contribution to the affordances of social media posts. This process led us to finalize a spectrum of features that could work in various use cases. For instance, we hypothesized that maintaining a consistent language style among a user's posts could enhance their self-presentation, contributing to the \textit{visibility} affordance. In contrast, tailoring a post to be published within a small team would benefit the users' association, as it could strengthen their existing social ties, aligning with the \textit{association} affordance. We identified a set of eight features that could help users adaptively tailor their posts for various workplace use cases, aligned with different affordances. We then used participants' feedback to identify the most essential functions that should be implemented as features in the tool. We also remained attentive to any new aspects or insights that participants might raise, which were not covered by the existing literature. To avoid limiting users' inquiries to specific features, we started by asking participants to imagine what features an LLM-empowered writing assistance system should have, before presenting them with our proposed features. 

For our second objective, we designed and implemented a low-fidelity prototype of the Corporate Communication Companion for users to explore and test its usability. To gather feedback, we used questions developed from the Technology Acceptance Model (TAM)~\cite{marangunic2015technology,benamati2002application,mun2003predicting}. Specifically, participants were asked whether they perceived the personalized communication system to be useful and whether they found it easy to use.

In practice, an interview session began with general questions about participants' experience with workspace social media posting (e.g., how often do they use the workplace social media platform, what are the most common topics they post about, how they expect their friends and colleagues to react to their posts). Next, we systematically gathered people's opinions on the two sets of questions about their independent writing process, and their preferences for the proposed features, including any additional new features that they suggested. After that, we introduced the prototype of the Corporate Communication Companion system, allowing participants to freely explore its functionalities and features. We actively encouraged them to share their thoughts on the usability of the system. 

In total, we recruited 7 active workplace social media users from a large corporation as our participants. The participants consisting of 2 females and 5 males, aged from 25 to 55, came from a variety of professional backgrounds in the corporation (e.g., tech lead, product manager, intern)\footnote{For detailed demographic information, please refer to the supplementary material.}. The interview sessions took 60 minutes in average, and each participant received a \$25 gift card for compensation. Recordings were taken and transcribed using Microsoft Teams. The authors conducted thematic analysis via inductive coding on participants’ responses to identify themes \cite{braun2012thematic}, which were subsequently refined through discussions across multiple sessions.


\subsection{Design Goals}
To address the challenges and needs identified from the interviews, we developed the following design goals for our system.

\xhdr{Design Goal 1: Recognize Blended Use Cases.} As highlighted in previous studies (\cref{workplace_social_media}), it is clear that distinct use cases coexist on the platform.
Our interview results further confirmed these findings. The primary use cases mentioned by participants were coded into two themes\footnote{Note that we excluded a third class of posts from our study scope, where people seek help with practical issues, such as corporate policy inquiries or technical issues related to their work devices. These posts are more utilitarian and less relevant to the personal image shaping within the workplace social media context. They also do not require extensive tailoring since they tend to be highly conversational and informal in nature.}: (1) Work-related, professional posts such as summarizing recent work, sharing research progress, and advertising projects, and (2) Leisure posts like book reviews, entertaining content generated by AI, humorous movie quotes, and experiences with interns. 

As a result, we begin with an requirement for the writing assistant to be aware of different use cases. Posting on workplace social media involves stakes. As P1 mentioned, ``\textit{there are some audiences [whom I am posting to], you know, I can be goofy,}'', ``\textit{Some messages could go wide and far. They are really meant as a broadcast,}'' suggesting the need for use case awareness and targeted strategies for each scenario. Assigning an undesirable attribute to a post, such as making a casual post overly formal or using inappropriate language in a professional post, can have undermine the affordances of visibility and association, respectively. Therefore, the writing assistant should infer the user's intent and adapt its strategy to provide help according to the use case.



\xhdr{Design Goal 2: Provide Post Outline at Structural Level.} Participants shared that outlining the post before writing could help them organize and process information. For instance, P7 said ``\textit{I tried to summarize it [the research to share] in like 5 to 10 bullet points}.'' 
This outlining strategy appeared common among participants in writing workplace social media posts. 
P5 shared that outlining is beneficial because ``\textit{it gives the clarity to the thought process}'' and ``\textit{it assists you to keep track of something}.'' 
However, we also found that some participants favored a more conversational style of communication. P1 shared that he ``\textit{view[s] [workplace social media writing] as conversations.}'' Between P1 and P5, P6 indicated that he writes ``\textit{to reach a balance between heuristics in structure and flow and analytical thinking in points to share with audience}.'' 
Therefore, an outline should be context-aware, adaptable to personal preferences in the writing process, and serve to support creativity. 

\xhdr{Design Goal 3: Support Post Editing at Language Level.} Participants demonstrated various strategies for editing their posts at the language level to suit their goals and intended audience. In general, participants favored a professional and formal tone, especially for professional posts. However, some individuals also wanted to infuse some fun into their work-related posts. For instance, P4 mentioned that she often uses cute emojis in research advertisements to make them engaging and align with her research community's norms (which leads to the benefits on the visibility and the association affordances). Similarly, participants emphasized the granularity of the post. For instance, P3 said ``\textit{being concise is my style},'' while P7 shared that he wanted posts to be more detailed via ``\textit{getting access to internal documents}.'' These preferences may seem contradictory, but both personal styles enhance visibility and persistence on workplace social media as they contribute to a consistent personal brand over time. 

Our interviews also highlighted some limitations of using standard writing assistance tools, such as ChatGPT, to craft workplace social media posts to achieve specific affordances. For instance, some participants expressed dissatisfaction with the style of content generated by standard writing assistants. Both P1 and P2 opined that content generated by this type of writing assistants ``is too vague.'' 
Moreover, the absence of personal information and context in such standard writing assistants poses additional challenges, as they may affect the accuracy of generated content.
Hence, our system aims to provide assistance at the editing level, allowing users to tailor their posts to a specific style that matches with their preferences and goals. 



\xhdr{Design Goal 4: Enable Customization Adaptive to Various Use Cases.} In line with the principles of mixed-initiative design, our goal is to strike a balance between providing assistance based on users' needs and allowing them to have effective control over the collaborative writing process. 
The current standard approach to providing writing assistance often adopts a laissez-faire interaction mode, where users are expected to specify every single requirement themselves.

However, this freestyle interaction can place a significant burden on users, as it is both time-consuming and prone to failure. Our research findings revealed that participants often neglected to provide context for their input queries, potentially due to the cognitive biases that prevented them from realizing how context-dependent some workplace communications are. An illustrative example shared by P2 highlighted this issue, where ChatGPT misinterpreted her input ``space evolution'' as related to ``cosmos'' and astronomy, whereas her actual intent was to discuss the ``space evolution'' of office buildings.
This misunderstanding necessitated her to invest extra effort into correcting the miscommunication. Consequently, interactions with writing assistants can suffer from a cold-start problem, where users must learn how to formulate natural language queries as prompts to steer the model's behavior. This can be especially demanding for lightweight tasks like workplace social media writing, where the benefit may not outweigh the effort required to master prompt engineering.


Therefore, we propose a hybrid approach that enables the system to take on part of the communication between users and the LLM. This approach recognizes that users may not have a clear mental model of how to interact with LLM-based tools. 
By providing the writing assistant with contextual information and allowing it to infer users' intentions in a complementary way, our system aims to reduce the burden on users. This way, users do not have to repeatedly and deliberately resolve potential issues that may arise from either insufficient contextual information or unclear user queries.

\subsection{Design and Implementation}

\begin{figure*}[t]
\centering
\includegraphics[width=1.0\textwidth]{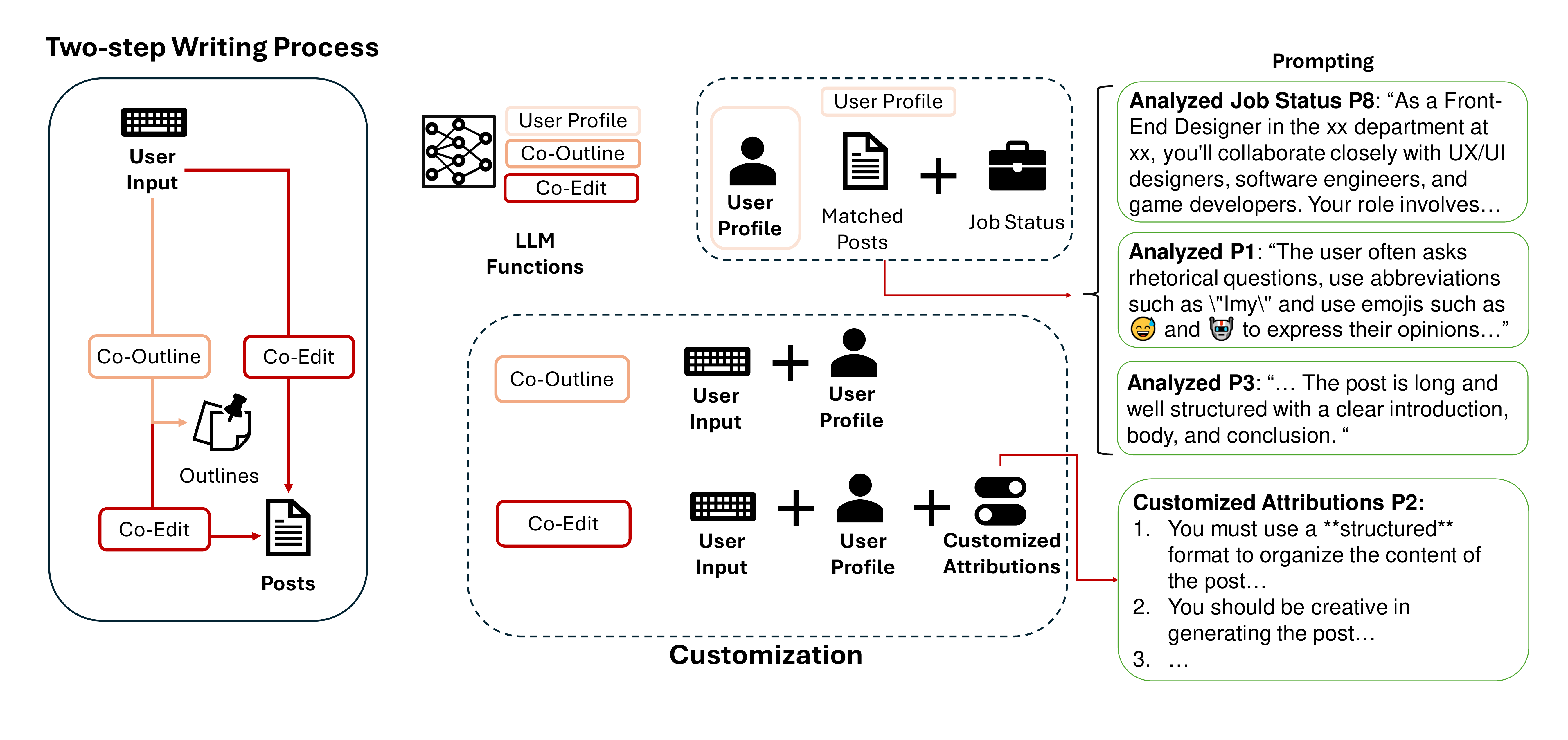}
\vspace{-10pt}
\caption{An illustration of the CCC system. The LLM-empowered system (1) \textbf{matches} the user input with its relevant historical posts that share the same use case as the input (i.e., professional or casual post). (2) establish \textbf{user profiles} based on job status and relevant historical posts (3) help users \textbf{Co-Outline} and \textbf{Co-Edit} based on user input with user profile and customization taken into account. The examples show (1) the LLM-summarized job status and writing styles of participants from their historical data (2) and customized attributions set by participants. This information was utilized by CCC in prompting for enhancing the system's writing assistance.}
~\label{fig:exp1:system}
\vspace{-10pt}
\end{figure*}

\begin{figure*}[t]
\centering
\includegraphics[width=0.8\textwidth]{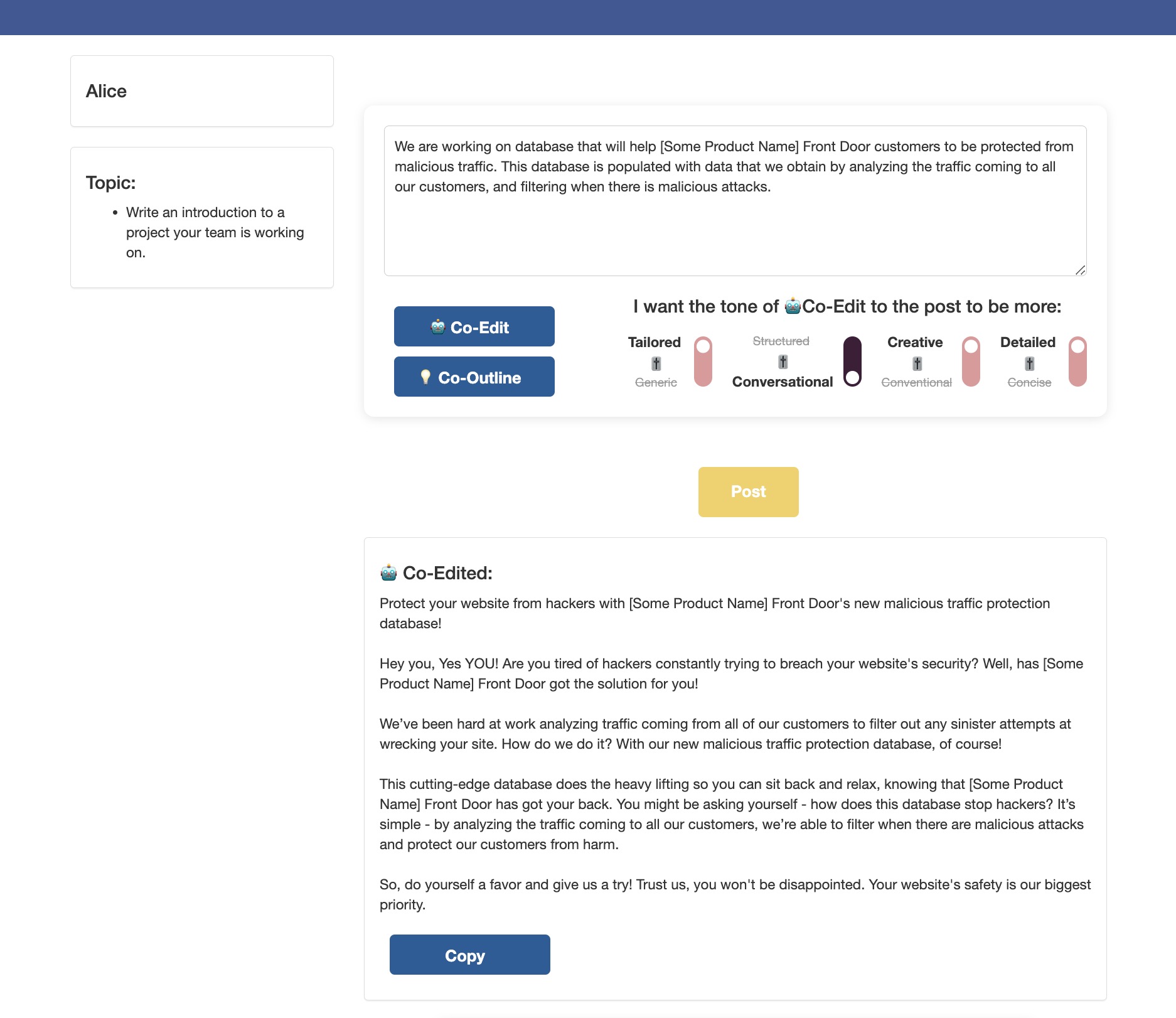}
\vspace{-10pt}
\caption{An illustration of the interface. The example shows Alice writing an introduction to a work about some security product that her team is working on. She decides to make the tone of Co-Edit to be more \textit{(tailored, conversational, creative, detailed)}.}
~\label{fig:exp1:interface}
\vspace{-10pt}
\end{figure*}

In general, CCC serves as a text editor that integrates the post-composition area on the user interface (\cref{fig:exp1:interface}). It takes any keywords, instructions, and drafts from the text input area and generates content based on user requests. The backend server is written in Python Flask and empowered by the OpenAI API.

To begin with, we designed the following functions for CCC to help people write workplace social media posts. An illustration of our proposed CCC system is provided in \cref{fig:exp1:system}. Hereby, we use Alice's one of the post-writing experiences facilitated by CCC as an example to describe the features of CCC in detail.

\xhdr{Use Case Matching.} We first employed the LLM to predict the intended use cases from users' historical posts. Specifically, these posts were classified into either ``professional'' or ``casual'' categories by leveraging the LLM as a classifier. The same classifier was also applied to recognize the intended use case from the user input. By doing so, we could identify the historical posts that matched the use case of the input, and consider them as relevant historical posts for the input. For instance, if Alice typed ``Make a post about our team picnic last Friday'', then her input would be classified as ``casual'', and her previous casual posts would be matched as relevant historical posts.

\xhdr{User Profile Creation.} 
After retrieving relevant historical posts that matched to the user input, we are able to adaptively establish the user profile in response to different use cases. In total, we gathered two types of information. First, we used the relevant historical posts to probe the users' personal writing voice and tone. Specifically, we prompted the LLM to summarize the main points and the writing sytles of each user's matched posts from various aspects, including but not limited to sentiment, wording, usage of emojis, etc. For the detailed prompts used, please refer to the supplementary material. This resulted in a set of outline summaries and writing style descriptions of the user's relevant historical posts. Secondly, we used users' public job status, including their job title and the organization within the corporation to which they belong (e.g., machine learning scientist in the device development department), to generate a brief job description of the user. For instance, Alice's profile for her casual topics would include she is a \texttt{machine learning scientist} who often follows \texttt{an order of introduction, experience sharing, and call for action}, and uses lots of \texttt{emojis}. In this way, users can have different profiles depending on the relevant historical posts matched, which enables the profile creation to adapt to various use cases and personal information at the same time.


These two features were used to extract the user intent and the contextual information of the post to be drafted. In particular, the generated user profile was used as a ``middleware'' component that enhances the following prompts with a prompt augmentation strategy (similar to \cite{wei2022chain}). This strategy aims to elicit the model's reasoning capacity and generate posts that reflect the user's goal and preference. Next, we present the two main functions of our Corporate Communication Companion (CCC) writing assistance system.

\xhdr{Co-Outline.} For the Co-Outline function, we filled in a prompt template containing the user's input query and a subset of their user profile, consisting of their job status and \textit{outline summaries of the relevant historical posts}. The filled template was fed to the LLM, which was instructed to produce an outline of a workplace social media post, taking the user's input and the user profile into consideration.

\xhdr{Co-Edit.} For the Co-Edit function, we followed a similar approach, using a prompt template with the user input and a subset of the user profile, while consisting of users' job status and \textit{previous writing style} instead. Moreover, in Co-Edit, we introduced a \textit{customization} feature, which enabled users to have the agency to specify the tone and voice of the post by adjusting the toggles on the interface. Each toggle represents a pair of contrasting attributes for the tone, and users can choose from the following options:
\begin{itemize}
    \item \textbf{Tailored vs generic}: `Tailored' indicates that the post will mimic the user's previous writing style, while `generic' indicates that the post will avoid a personal writing style.
    \item \textbf{Conversational vs structured}: `Conversational' indicates a more casual tone, whereas `structured' indicates a more professional or formal tone.
    \item \textbf{Creative vs conventional}: `Creative' indicates a more imaginative writing style, whereas `conventional' indicates a more straightforward style.     \item \textbf{Detailed vs concise}: `Detailed' indicates that the post will contain more in-depth information, whereas `concise' indicates that the post will be shorter and to-the-point.

\end{itemize}
These toggles yield a total of 16 potential combinations of tones. For instance, a user could make the tone of their post co-edited by CCC to be (\textit{tailored, conversational, creative, detailed}). Each toggle contributes a predefined instruction to the prompt template. Furthermore, to ensure the generated content meets the user-specified instructions, we also took a multi-step prompting strategy. After producing a piece of content, we prompted the LLM again with a separate template to check whether the generated content fulfilled all the criteria specified by the user. If any violations were flagged, the content would be revised accordingly.

To illustrate the benefits of the proposed system, consider Alice's writing example again. She can then use the Co-Outline and Co-Edit functions to enhance her post-writing process in a flexible way. She can iteratively alternate between Co-Outline, Co-Edit, natural language interactions in the text box, and manual editing until she is satisfied with her post.

The proposed CCC system can meet the design goals by providing the abovementioned features. It recognizes the users' intended use case, then matches it with relevant historical posts and the ongoing user input, which supports \textbf{Design Goal 1}. It allows users to use the Co-Outline and Co-Edit functions at different writing stages, which supports \textbf{Design Goal 2} and \textbf{Design Goal 3}, respectively. Finally, it enables a balanced interaction mode between users and the system, by allowing users impart customization based on their preferences and letting the system infer the contextual information, thereby supporting \textbf{Design Goal 4}.

\section{Phase 2: How Do Users Like the Customized Communication Companion?}

To evaluate the effectiveness and usability of CCC, we conducted a within-subjects user study with 10 participants. 

\subsection{Participants}
We recruited a total of 10 participants through a combination of convenience sampling and direct recruitment. We invited potential participants via the company's mailing lists and directly contacted active users of the internal workplace social media platform in question, excluding participants who have already participated in our Phase 1 study. All participants possessed at least basic experience in using generative AI for writing assistance (e.g., ChatGPT) and composing content for workplace social media. 6 participants had actively posted on the internal workplace social media platform within the previous two months. The remaining 4 participants had experience on similar professional social media platforms such as LinkedIn and academic Twitter. The participants consisting of 4 females and 5 males, aged from 23 to 50, came from a variety of professional backgrounds in the corporation (e.g., product manager, researcher, engineer, intern). Each received a \$25 gift card for compensation. 

\subsection{Tasks}

We designed the following four writing tasks for the user study. Specifically, two writing tasks are work-related, professional topics:
\begin{itemize}
    \item \textit{Make an announcement of a talk or presentation that you will give.}
    \item \textit{Write an introduction to a project your team is working on.}
\end{itemize}
The other two tasks are related to leisure and casual topics:
\begin{itemize}
    \item \textit{Make a post about an intern event.}
    \item \textit{Make a post about a team-building event.}
\end{itemize}
These four writing tasks were developed based on both previous literature and the feedback from our interviews. We consider that they are representative of the most common use cases that people encounter on the company's internal workplace social media platform, and at the same time are use cases that could benefit from writing assistance.

\subsection{Protocol}

The user study consisted of two separate sessions conducted over two days: A post-writing session and a post-evaluation session. Prior to their participation, all participants provided their consent for the study and were requested to provide a total of four social media posts: Two work-related professional posts and two leisure-related casual posts that they had previously shared on their workplace social media accounts.

On Day 1, we conducted an in-person post-writing session where participants collaborated with a writing assistant to compose posts on given topics and shared their user experience. Each participant was instructed to complete the four previously mentioned writing tasks in each session. The four tasks were split into two sets, where each set contained one professional topic and one casual topic. Participants were asked to use both CCC and a standard writing assistant to complete each set of tasks. To mitigate any potential learning effects, the order of tasks and writing assistant assignments was randomized. Before attempting to use each system, participants were given a pre-written introduction describing the functions provided by the system. Additionally, participants were asked to complete a tutorial task to familiarize themselves with the system.
 
At the beginning of each task, we encouraged participants to reflect on their prior experiences or upcoming events related to the topic, serving as cognitive priming to facilitate writing. Participants then composed their posts and proceeded to the next task once satisfied with the outcome. After completing each task, participants were asked to complete a survey on their user experience. After completing all four tasks, participants were asked to provide feedback on both writing assistance systems. Each user study session was recorded with the participant's permission and took 40 minutes on average.

On Day 2, participants were asked to complete an online survey to evaluate posts written by other participants. Each participant reviewed four randomly sampled posts from a pool of posts composed by others. \footnote{We only asked participants to evaluate four posts primarily for their time convenience. Considering the limited number of participants, only a random half of the posts we collected on Day 1 were evaluated on Day 2.}

Recordings were taken and transcribed using Microsoft Teams. The first author conducted thematic analysis via inductive coding on participants' responses to identify themes, which were subsequently refined through discussions with co-authors across multiple sessions.



\subsection{Measures and Analysis Methods}

In the Day 1 survey, we evaluated the effectiveness of the proposed system with a focus primarily on the \textbf{writer}'s perspective. Specifically, we divided the questionnaire into two main parts: (1) questions about participants' writing experience and cognitive load and (2) questions regarding participants' attitudes toward the writing assistant. 

The first set of questions was crafted with the understanding that a positive writing experience is essential for users' intrinsic motivations to participate in online communities. Starting from the , we first used metrics to understand participants' perceived assistance from the writing assistant via the ratings on their \textbf{Collaboration}: \textit{"I felt like I was collaborating with the system."} and \textbf{Engagement}: \textit{"Using the writing assistant feels engaging."} with the writing assistant, as well as their perceived attribution in the writing process in \textbf{Attribution}: \textit{"The post written using the writing assistant was totally due to my attribution."} and \textbf{Authenicity}: \textit{"I feel the audience will consider that the post is written by myself independently."}. In addition to metrics probing into the writing process, we would like to further gauge participants' views on their written posts in terms of the \textbf{Completeness}: \textit{"The post I wrote using the writing assistant feels complete (e.g., there’s nothing to be further worked on)."} and \textbf{Uniqueness}: \textit{"The post I wrote using the writing assistant feels unique."}. Additionally, we included an overall evaluation of their experience in \textbf{Satisfaction}: \textit{"I am satisfied with the writing process."}. All items were rated on a 7-point Likert scale, ranging from 1 (Strongly disagree) to 7 (Strongly agree). 

Furthermore, to assess participants' cognitive load in writing, they answered the effort questions rephrased from the NASA-TLX question set in the context of using the writing assistant. Similarly, this set of questions also used the 7-point Likert scale, with 1 indicating very low and 7 representing very high. 

Beyond the perceived quality of writing process and outcome, attitudes towards the intelligent system play a crucial role in shaping people's current utilization and future adoption of the systems. Hence, motivated by the growing line of research in the broader human-AI collaboration spectrum \cite{jacovi2021formalizing,lai2021towards}, the second set of questions focuses on people's attitudes to AI. More specifically, we employed the metrics based on the expanding human-AI interaction paradigm \cite{yin2019understanding,zhang2020effect}. The metrics incorporated the ratings on participants' perceptions in several crucial dimensions including
the \textbf{competence} of (\textit{"The content that the assistant provides to me is as good as that which a highly competent person could provide."}), \textbf{faith} in (\textit{"The content that the writing assistant provides to me is as good as that which a highly competent person could provide."}, \textbf{willingness} to use (\textit{"I will be willing to use the writing assistant in the future."}, and overall \textbf{trust} (\textit{"Overall, I trust in the writing assistant."}) in the writing assistant.

In the Day 2 survey, our focus shifted to assessing the perceived quality of the posts from the \textbf{audience}'s perspective. Recognizing that online communities are not just about the writers but also the readers, we understand that high-quality compositions can significantly enhance community engagement and interaction \cite{otterbacher2009helpfulness,fan2022quantifying}. This is particularly relevant in our context of workplace social media, where the impact of posts extends beyond individual expression to the influence in a broader community. Therefore, we would like to understand how these posts composed alongside the writing assistants were perceived in terms of being \textbf{Informative}, \textbf{Engaging}, \textbf{Appropriate}, and of a high overall \textbf{Quality}. Similarly, we asked participants to rate their level of agreement on statements following the structure like \textit{"I think the post is informative/is engaging/is appropriate to publish on the platform/overall has a high quality."}, on a 7-point Likert scale.

To analyze the quantitative results obtained from surveys, we used the paired T-test to examine the potential differences between participants' ratings between treatments with assistance from different writing assistants. For the qualitative feedback that we collected during the user study, similarly, we conducted the thematic analysis via inductive coding on the transcripts recorded in the user study.




\begin{figure*}[t]
     \centering

      \begin{subfigure}[b]{0.7\textwidth}
         \centering
     \includegraphics[width=\textwidth]{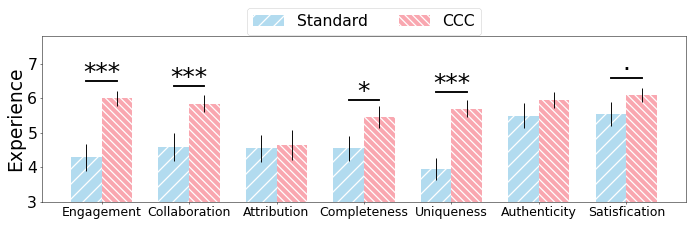}
     \caption{User Experience}
     \label{fig:all_tasks:experience}
     \end{subfigure}

     \begin{subfigure}[b]{0.49\textwidth}
         \centering
     \includegraphics[width=\textwidth]{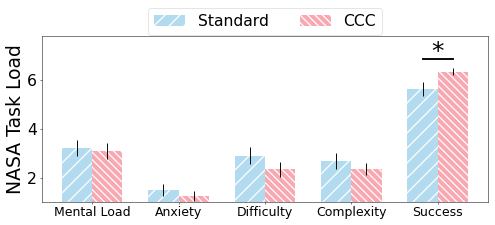}
     \caption{NASA Task Load}
     \label{fig:all_tasks:load}
     \end{subfigure}
          \centering
     \begin{subfigure}[b]{0.39\textwidth}
         \centering
     \includegraphics[width=\textwidth]{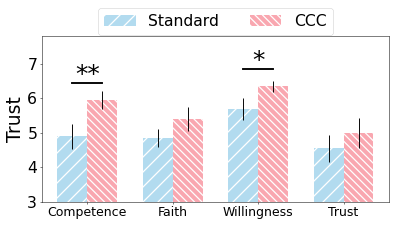}
         \caption{Trust}
     \label{fig:all_tasks:trust}
     \end{subfigure}
     \vspace{-5pt}
    \caption{Participants' self-reported user experience (\cref{fig:all_tasks:experience}), perceived task load (\cref{fig:all_tasks:load}), and trust (\cref{fig:all_tasks:trust}) in using CCC and standard writing assistance. Results show that people felt the writing experience with CCC is more engaging and collaborative, and the posts written with CCC are more unique and complete. As a result, people perceived that CCC is more competent, and they are more willing to use CCC. Such enhancement did not cause extra cognitive loads.
    Error bars represent the standard errors of the mean.
    }
        \label{fig:all_tasks:all_measurements}
     \vspace{-10pt}
\end{figure*}

\begin{figure*}[t]
     \centering

      \begin{subfigure}[b]{0.4\textwidth}
         \centering
     \includegraphics[width=\textwidth]{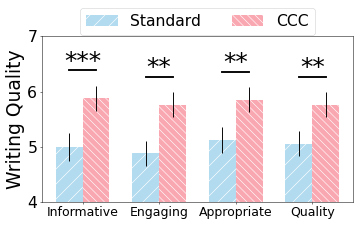}
     \caption{Writing Quality}
     \end{subfigure}
     
     \vspace{-5pt}
    \caption{Participants' evaluation of posts composed by other participants using CCC and standard writing assistance. People felt the posts written by others under the help of CCC were more informative, engaging, appropriate, and thus overall in a better quality.
    Error bars represent the standard errors of the mean.
    }
        \label{fig:all_tasks:evaluation}
     \vspace{-10pt}
\end{figure*}

\subsection{Results}
\subsubsection{Writers: CCC Enhances Users' Writing Experience} 

To understand how users' writing experience differed when using CCC versus the standard writing assistant, We analyzed the recordings and the post-task feedback survey from the \textbf{writer}'s perspective. \cref{fig:all_tasks:experience} compares the feedback from participants on the abovementioned four writing tasks in terms of \textbf{interactivity} (i.e., Engagement, Collaboration), \textbf{perceived quality} (i.e., Completeness, Uniqueness), \textbf{perceived ownership} (i.e., Attribution, Authenticity) and overall \textbf{satisfaction} of their post writing experience. 
We found that participants rated interactivity with CCC as higher than that of the standard writing assistant. Results from the paired T-test suggest that participants felt CCC was more engaging and collaborative than standard writing assistance (Engaging - CCC: $\mu = 6.00, \sigma = 1.02$, Standard: $\mu = 4.30, \sigma = 1.75, t(19) = 5.84, p < 0.001 $; Collaborating - CCC: $\mu = 5.85, \sigma = 1.08$, Standard: $\mu = 4.60, \sigma = 1.82, t(19) = 4.07, p < 0.001 $). Next, we considered participants' own perceptions of the posts. With the aid of CCC, participants felt that their final posts were more complete and unique. Again, the paired T-test confirmed the significance of the difference (Unique - CCC: $\mu = 5.45, \sigma = 1.08$, Standard: $\mu = 4.55, \sigma = 1.39, t(19) = 5.04, p < 0.001 $; Complete - CCC: $\mu = 5.85, \sigma = 1.08$, Standard: $\mu = 4.60, \sigma = 1.82, t(19) = 2.31, p = 0.032 $). However, there was no significant difference in participants' sense of of ownership over their posts. As a result, participants were slightly more satisfied with CCC ($\mu = 6.10, \sigma = 0.91$) than with the standard writing assistant ($\mu = 5.55, \sigma = 1.61, t(19) = 1.81, p = 0.086$), but this difference was not statistically significant.

Regarding ease of use, \cref{fig:all_tasks:load} illustrates the NASA task load scores for participants during their writing tasks. Participants generally rated CCC as slightly easier to use, and felt more successful in writing with CCC than with the standard writing assistant (CCC: $\mu =  6.35 ,\sigma= 0.67 $; Standard: $\mu =  5.65 ,\sigma= 1.27 $; $t(33) =  2.77 , p = 0.012$). We did not find statistically significant differences in other aspects of task load, such as mental load, anxious feelings, perception of difficulty, and complexity in using the system. However, there was a consistent trend of lower task load scores for CCC across these dimensions.

Consequently, users' improved writing experiences with CCC, without imposing an additional cognitive load, contributed to a consistent trend in \cref{fig:all_tasks:trust}, where users expressed higher trust~\cite{mayer1995integrative,glikson2020human,yin2019understanding} in CCC. Specifically, paired T-tests confirmed that participants considered CCC as more competent ($\mu =  5.95 ,\sigma= 1.19 $) than the standard assistant ($\mu =  4.9, \sigma= 1.59 , t(33) = 3.8, p = 0.0012$) and indicated a greater willingness to use CCC ($\mu =  6.35 ,\sigma= 0.67 $) in the future compared to the standard assistant ($\mu =  5.7 ,\sigma= 1.45 $; $t(33) =  2.46 , p = 0.024$).

Together, our findings demonstrate that CCC is an easy-to-use and user-friendly interactive system that significantly improves the writing experience. Posts created with the assistance of CCC are more unique and complete from the writer's perspective. Therefore, users perceive CCC as competent and show a higher preference for it.

\subsubsection{Audience: CCC Improves Users' Writing Quality} At the same time, the evaluation survey from Day 2 of our user study enables us to verify the effectiveness of CCC from the \textbf{audience}'s perspective. We received evaluations from 9 participants on 68 posts that were written in the user study\footnote{One participant dropped out from the survey due to personal issues, and one participant left office after Day 1.}. \cref{fig:all_tasks:evaluation} compares the participants' ratings of the posts composed by others with the aid of CCC and the standard writing assistant. The results show a consistent trend that audience members found the posts co-created with CCC as more informative (CCC: $\mu =  5.88, \sigma= 1.32 $; Standard: $\mu =  5.0, \sigma= 1.48 $; $t(33) =  2.93, p < 0.001$), and they were more willing to engage with these posts than those co-created with the standard writing assistant (CCC: $\mu =  5.76, \sigma= 1.35 $; Standard: $\mu =  4.88, \sigma= 1.34 $; $t(33) =  2.93, p = 0.0018$). As a result, they rated the posts written with CCC as more appropriate for publishing on the internal workplace social media according to the intention of the post (CCC: $\mu =  5.85, \sigma= 1.31 $; Standard: $\mu =  5.12, \sigma= 1.37 $; $t(33) =  2.93, p = 0.0025$). Ultimately, the evaluation confirmed that CCC improved participants' writing quality from the viewpoint of potential audience members on workplace social media platforms (CCC: $\mu =  5.76 , \sigma:= 1.28 $; Standard: $\mu =  5.06 , \sigma= 1.35 $; $t(33) =  2.93 , p = 0.0061$).

\subsection{Exploratory Observation of Participants' Behaviors}


     


\begin{figure}[tbp]
    \centering
    \begin{minipage}{0.5\textwidth}
        \centering
        \includegraphics[width=\linewidth]{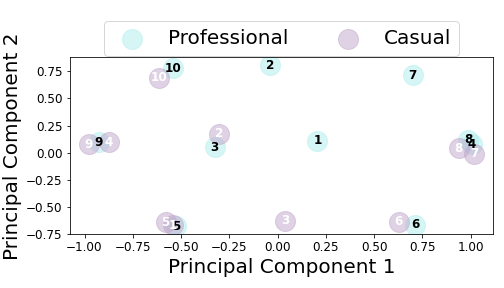}
        \caption{Scatter plots of participants' customization settings for the tone of Co-Edit. Numbers alongside dots indicate the index of the corresponding participant from 1 to 10.}
        \label{fig:exp:customization}
    \end{minipage}%
    \hspace{20px}
    \begin{minipage}{0.43\textwidth}
        \centering
        \hspace{20px}
        \includegraphics[width=\linewidth]{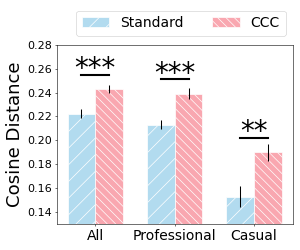}
        \caption{Averaged pairwise cosine distance between text embeddings of posts written by participants using the standard writing assistance or CCC, on all tasks, professional tasks, and casual tasks, respectively.}
        \label{fig:exp:content}
    \end{minipage}
            \vspace{-10px}
\end{figure}

\xhdr{Distribution of Customization.} \cref{fig:exp:customization} displays a scatter plot depicting the participants' final customization settings. 
To obtain vector representations of participants' customization settings, we encoded each pair of opposing attributes as binary values (0 or 1), resulting in a four-dimensional vector. We then applied principal component analysis (PCA) to reduce the dimensionality to two, allowing us to visualize these customizations. The plot reveals that participants adopted very different strategies to adjust the tone of their posts with the Co-Edit function (i.e., we didn't observe a clear separation between professional and casual posts). This observation strongly supports the assumption that individuals adopt varying tailoring preferences for their workplace social media posts, even for the same use case. Moreover, it aligns with our findings that people perceived the posts they composed as unique, as they requested edits in different ways to suit their specific preferences.

\xhdr{Distribution of Produced Content.} As we highlighted in previous sections, users often employ diverse strategies and exhibit different preferences in creating workplace social media content. This diversity is an important factor for LLM writing support. If the writing support does not offer a broad range of options, users may feel dissatisfied with the writing process and produce posts that lack originality and variety within the community.
To investigate how CCC affects the diversity of the produced posts, we conducted an analysis on the post content written by participants with CCC. We utilized the pre-trained OpenAI text-embedding-ada-002 model to generate embeddings for each post authored by participants. Subsequently, we calculated the cosine distance between every pair of posts for all writing tasks, professional tasks, and casual tasks separately. We then used this distance as a metric to quantify the diversity within a group of posts. This allowed us to compare the posts written with the standard writing assistant and those written with CCC. \cref{fig:exp:content} presents the results of averaged cosine distances among posts. We observed a consistent trend of higher diversity among pairs of texts composed using CCC. A set of Mann-Whitney U-tests on the cosine distances among posts written with CCC and the standard writing assistant confirmed the significance of the differences, suggesting posts written with CCC were significantly more diverse in all use cases ($p < 0.001$), in professional posts ($p < 0.001$), and in casual posts ($p = 0.0022$). 

Contrary to the intuitive expectation that unconstrained natural language interactions in writing assistance would naturally lead to more diverse compositions. However, our observations indicate that people often produce more homogenous posts instead. There are several plausibe reasons for this. Firstly, it can be challenging for individuals to come up with original ideas for their workplace social media writing. Additionally, the blending of thinking and writing processes could encourage heuristic writing. In such cases, individuals might feel compelled to include the first or easiest idea that comes to mind, rather than exploring more unique or customized composition. As a result, rather than expanding the range of expression, unrestricted natural language interaction might narrow it, leading to more uniform and content.

\subsection{User Preference and Feedback}

At the end of Day 1 in the user study, we asked participants to provide feedback as a qualitative probe into their experience using CCC compared to the standard writing assistance. We coded participants’ feedback to this question and identified three different themes.

First, participants found that CCC allowed for more efficient control over the content generation process because of its configurability. Participants suggested that natural language instructions specifying their customization needs were more difficult to compose and control.
Indeed, three participants mentioned that their natural language queries failed to capture their specific requirements for tailoring the post. For instance, P7  said that ``\textit{the second one [the standard writing assistance system] is hard to control. I don't think it understands my requests.}'' Conversely, for users with LLM expertise, natural language queries ``\textit{could also be redundant},'' as noted by P8. In some cases, their expertise further encouraged them to uncover the system's mechanism, leading them to put extra mental effort into exploration. P1, as a machine learning scientist, said ``\textit{I will keep asking, what's wrong with my prompts in version one [the standard writing assistant]}''. With the customized attribution settings, this process became more efficient. Similarly, P9 shared that \textit{``the toggles really help with saying immediately like that's way too long for something that I would post}.'' 
Likewise, the perceived transparency of the content generation process was enhanced due to the customization settings--``\textit{otherwise it is a black box},'' noted P8. 

Participants also found Co-Editing via CCC was more effective in developing a specific writing style that could be tailored to match their individual writing preferences. This style encompassed both the structural flow of composition and the choice of words. For example, P3 stated, ``\textit{These two sentences are definitely my style, I don't know but it just sounds like me.}'' At the wording usage level, three participants mentioned that their exact linguistic features (e.g., abbreviations like ``y'all'', exclamations, emojis) were well-captured by CCC. In addition, CCC helped people in obtaining a specific writing style for their corresponding use case more efficiently. 

The decomposition of the writing process was found to be helpful as well. Three participants mentioned that Co-Outline helped them to begin writing their posts and to transfer ideas from drafts. Participants confirmed the ideas generated by the Co-Outline function were useful and adopted them into their editing. Four participants chose to write posts by following the bullet points in the generated outline one by one. P4 commented, ``\textit{The outline was actually very helpful in coming up with how I should structure things and how to even just get started. I think that's pretty challenging.}'' 
P9 specifically indicated a use case ``\textit{for users who want more structure in their posts}.'' Furthermore, the incorporation of the user's job status into the outline enabled the generation of contextually relevant content. For instance, P1, who is the lead of a machine learning team, expressed appreciation for the CCC-generated idea of organizing a hackathon or an AI-based game contest.

Participants also pointed out several limitations in the current version of CCC. Some participants expressed a desire for enhanced customization options. Two participants suggested allowing CCC to access their previous work for deeper tailoring and enabling posts to incorporate more technical details. Additionally, one participant advocated for finer granularity in customizing attributions. Some participants also offered feedback from the usability perspective. Two participants mentioned the utility of providing access to the edit history for reference across different editing trials, in addition to the existing comparison between the input and output areas. Furthermore, one participant noted the absence of multi-modality in editing, as images are often included in their workplace social media posts.

\section{Discussion}

In this paper, motivated by the observed complex use cases and various personal differences of workplace social media writing, we present the design, implementation, and evaluation of a novel LLM-empowered writing assistant. The system provides tailored, customized, contextual assistance throughout the workplace social media post writing process and helps users improve their professional images for different use cases. In this section, we provide the implications and limitations of our work.


\subsection{Tailoring Content to Personal Preference} 

Workplace social media is a type of corporate communication that balances both formality and informality, i.e., it is neither as rigid as formal documents such as presentations or reports, nor as casual as online chat. Instead, workplace social media covers a wide spectrum of complex use cases, which require different levels of professionalism and different tones or styles. Our offline data analysis again confirmed the understanding. In this context, our results clearly demonstrate the benefits of leveraging individual preferences to tailor content generation. Such tailoring enables users to craft their posts more effectively and satisfactorily, thereby promoting increased content production on the workplace social media platform. Additionally, it encourages users, as consumers of content, to engage more actively with the posts due to their improved quality. These outcomes help foster a more thriving environment within workplace social media, where employees communicate with fewer obstacles and more effectiveness. Similarly, such tailoring could be extended to other communication scenarios in corporate settings beyond workplace social media.

It is important to note that our preference-based content tailoring is closely related but different from the concept of `personalization' in the previous literatures, which often assumes the persona reflected in the content matches the user's past behavior. However, there may be a discrepancy between the users' own persona and the persona projected via their generated content, where people may try to be, for instance, more professional or more humorous on a public communication space than they actually are. The customized attributes setting in CCC allows users to craft a specific tone that they prefer under that scenario, but that may differ from their actual persona. Therefore, such `personalization' should be implemented within a mixed-initiative interaction framework. In other words, the intelligent assistance offered to facilitate people's communication should respect users' agency and allow them to adjust, explore, and finalize the desired `persona' for each specific use case. This requires a deeper understanding of individual behavior in corporate communication.

Moreover, we observe that the degree of `personal style' should be appropriate for the context, depending on both the use case and individual preferences. For instance, professional and casual communication require different amount of personalization. At the same time, as mentioned by one of the participants, although CCC successfully captured her writing style, she was a little cautious about how much personal flavor should be incorporated into her posts.

\subsection{Human-LLM Collaboration: Interaction Design}
Our research suggests that access to an LLM alone is not enough to perform certain writing tasks such as composing workplace social media posts. Although all participants were familiar with ChatGPT-like products in their daily lives, they did not demonstrate clear mental models of how to effectively collaborate with LLM-based tools. There is insufficient evidence to support that they are aware of how to communicate their intentions to the LLM in natural language and obtain the desired results, even when the LLM is technically capable of fulfilling their tasks. In our user study, both systems utilized the same LLM API and base prompt format as the system setting, so users of the standard writing assistance are able to achieve similar or even identical outputs as CCC by modifying their input text. However, we observed discernible quality differences between the two treatments, debunking the assumption that a laissez-faire interaction style, where users and LLMs exchange requests in natural language without constraints, fully harnesses the potential of LLMs for these specific writing tasks. 

Consequently, in the era of LLM emergence, it is still crucial to thoughtfully design LLM-based tools with functions and interfaces that match user needs, rather than solely relying on LLMs and expecting them to satisfy all user demands. For instance, developing finer-grained collaboration workflows and feedback mechanisms between human and LLM writing assistance holds great promise, as we showed in our study as an example. To achieve this goal, we need to address this challenge from both user and LLM system perspectives. On the one hand, CCC demonstrates a regulated cognitive process that helps users structure their writing into separate stages and engage LLMs to assist at each step. This design compels both users and LLMs to tackle writing challenges in a step-by-step manner, which is a common problem-solving strategy~\cite{seggar1972conversion,croskerry2003cognitive, buccinca2021trust,becker2006review}. Another strategy that can improve human-AI collaboration is user education~\cite{yildirim2023investigating,chiang2022exploring}, which can help people develop a better mental model of how to use LLMs effectively. On the other hand, LLM applications should also be more responsive to users' behavioral patterns. This includes providing more relevant feedback and interactive interfaces that enable users to leverage the full potential of LLMs in their collaborative endeavors. By addressing these multifaceted considerations, we can navigate the evolving landscape of human-LLM collaboration more effectively.

\subsection{Human-LLM Collaboration: LLM Adaptation}
The results of our study exemplify the incorporation of the user profile as contextual information that prompts LLMs to collaborate more effectively with users. First, by integrating historical posts and job status into the LLM prompts, our system can better understand each user's writing style, tone, and appropriate content themes for posts. Further, via recognition of use cases, we matched the writing task with the appropriate contextual information. Thus, it enables the LLM to generate outlines and edits that are not only contextually appropriate but also aligned with the user's unique voice and past content strategies. 

Though the AI communities have proposed various methods to provide contextual information to LLMs, which is relevant to our first step, it is still under-explored how we should design context-aware LLM assistants in practice. Particularly, the challenge lies in determining when and how to utilize relevant segments of context within prompts. Thus, our second step enhances the system for better tailoring content via the matching between appropriate historical data with the current use case through user input (i.e., professional historical posts matched to professional use case). Such matching could be further enhanced via advanced algorithmic approaches, such as information retrieval from users' historical data. Similarly, our design could also serve as an example of user-centric LLM adaptation, which shows a comprehensive procedure from data analysis, an expert-involved use case matching, and the further integration of context information. Similar methodologies can be extended to various other scenarios. For instance, adapting LLMs to reflect language usage variations across different demographic features like age groups \cite{klein1996language}, personalities \cite{lee2007relations}, or varying use cases such as urgent care versus routine medical diagnosis in healthcare settings \cite{islam2016speaking}. 

Moreover, the field of sociolinguistics \cite{wardhaugh2021introduction} has pointed out the diversity of language preferences in different scenarios, partially inspiring the use case recognition for context provision in our study. Previous literature in this domain highlights the communication strategies across different social and cultural contexts \cite{roberts2010language}. Therefore, it would be beneficial for downstream applications involving human-LLM collaboration to draw from sociolinguistic knowledge. This would aid in understanding the motivations behind user inputs and allow the LLM to adapt to use cases without the need for manual adjustment of human input. Such adaptation could be implemented without fine-tuning instead via a system-set prompt as we did for CCC. By integrating these insights into LLM development, we can enhance the model's ability to understand and adapt to the varied linguistic and contextual nuances in various scenarios for user-friendly and effective collaboration.




\subsection{Limitations and Future Work}
Our study was conducted on a proxy system to one specific format of workplace social media (i.e., social networking platform), and involved one pair of specific user behaviors (i.e., post writing and consuming) around a fixed set of topics. Cautions should be used when generalizing results in this work to other settings, such as forum-based workplace social media, communication that involves multiple modalities (e.g., images, voice, videos), conversational communication, and other user behaviors, including commenting, forwarding, confirmation, etc. 
There are also specifications of the system that may limit the generalizability of our results. Our LLM prompts were empirically developed, without being tested on large-scale datasets due to the sparsity of research in this domain. We implemented the writing assistant as an editor, while other formats like chatbot can also be taken into consideration. Therefore, we encourage more research on how LLMs can enhance different types of communication, especially in professional settings. 


\section{Conclusion}
In this paper, we presented a two-phase study in light of the findings from the offline data analysis on a workplace social media platform. The study aims at designing and evaluating an LLM-powered writing assistant, CCC, specifically for workplace social media. In Phase 1, we interviewed potential users to explore their needs for LLM-based writing assistants on workplace social media. Based on the findings, we iteratively refined the design of our prototype. In Phase 2, we evaluated the effectiveness of CCC with a user study from both the writers' and audience members' perspectives. Our work makes unique contributions to the community in several ways. To start, it fills the gap in the existing literature on how users perceive and utilize intelligent systems for semi-formal corporate communication. Furthermore, it tackles the challenge by providing customized and controllable writing assistance at each stage of the writing process. Our study also provides insights on how to design user-friendly LLM-empowered tools for complex communication scenarios, especially in the context of workplace social media. 

\begin{acks}
To Robert, for the bagels and explaining CMYK and color spaces.
\end{acks}

\bibliographystyle{ACM-Reference-Format}
\bibliography{ccc}

\appendix

\end{document}